\documentclass[]{mn2e}
\usepackage{graphicx}

\newif\ifAMStwofonts

\def\gs{\mathrel{\raise0.35ex\hbox{$\scriptstyle >$}\kern-0.6em 
\lower0.40ex\hbox{$\scriptstyle \sim$}}}
\def\ls{\mathrel{\raise0.35ex\hbox{$\scriptstyle <$}\kern-0.6em 
\lower0.40ex\hbox{$\scriptstyle \sim$}}}

\newcommand{\be}{\begin{equation}}
\newcommand{\ee}{\end{equation}}
\newcommand{\bea}{\begin{eqnarray}}
\newcommand{\eea}{\end{eqnarray}}

\newcommand{\ud}{\mathrm{d}}

\title{Galaxy orbits and the intracluster gas temperature in clusters}
\author[Benatov et al.]{L. Benatov$^{1,2}$, K. Rines$^{2,3}$, 
P. Natarajan$^{1,2,3}$, A. Kravtsov$^{4,5}$ \& D. Nagai$^{4,5,6}$\\
$^1$Astronomy Department, Yale University, P.O. Box 208101, 
       New Haven, CT 06520-8101, USA \\
$^2$Department of Physics, Yale University, P. O. Box 208101,
New Haven, CT 06520-8101, USA\\
$^3$Yale Center of Astronomy and Astrophysics, Yale University, P.O. Box
208101, New Haven, CT 06520-8101, USA \\
$^4$Department of Astronomy and Astrophysics, University of Chicago, 
5640 S. Ellis Ave., Chicago, IL 60637, USA \\
$^5$Kavli Institute for Cosmological Physics, University of Chicago, 
5640 S. Ellis Ave., Chicago, IL 60637, USA \\ 
$^6$ Theoretical Astrophysics, California Institute of Technology,
Mail Code 130-33, Pasadena, CA 91125 \\ 
}

\begin{document}

\maketitle

\label{firstpage}

\begin{abstract}
In this paper we examine how well galaxies and intra-cluster gas trace
the gravitational potential of clusters. Utilizing mass profiles
derived from gravitational lensing and X-ray observations, coupled
with measured galaxy velocities, we solve for the velocity anisotropy
parameter $\beta_{\rm orb}(r)$ using the anisotropic Jeans equation.
This is done for five clusters, three at low redshift: A2199, A496 and
A576 and two at high redshifts: A2390 and MS1358. With X-ray
temperature profiles obtained from \textit{Chandra} and
\textit{ASCA/ROSAT} data, we estimate $\beta_X(r)$ the ratio of energy
in the galaxies compared to the X-ray gas. We find that none of these
clusters is strictly in hydro-static equilibrium. We compare the
properties of our sample with clusters that form in high-resolution
cosmological N-body simulations that include baryonic physics.
Simulations and data show considerable scatter in their $\beta_{\rm
orb}(r)$ and $\beta_X(r)$ profiles. We demonstrate the future
feasibility and potential for directly comparing the orbital structure
of clusters inferred from multi-wavelength observations with high
resolution simulated clusters.
\end{abstract}

\begin{keywords}

\end{keywords}

\section{Introduction}

Galaxy dynamics in clusters provided the first evidence of dark matter
(Zwicky 1933; 1937). However, the orbital distribution of galaxies
remains a major source of uncertainty in deriving mass profiles from
cluster galaxy dynamics. Several authors have shown that a wide
variety of mass profiles can reproduce the observed galaxy and
velocity distribution in clusters (The \& White 1986; Merritt 1987).
The observational errors in the measurements of velocity dispersion
profiles can translate to large uncertainties in dynamical mass
estimates. Even when the kinematic data available are extensive,
simplifying assumptions are usually necessary to estimate orbital
distributions from velocity data alone (van der Marel 2000; Biviano \&
Katgert 2004).  The key assumption in using galaxy kinematics to
derive the gravitational potential of the dark matter is that they are
robust tracers of the total mass distribution. However, as galaxies
represent only a small percentage ($\sim 1\%$) of the total mass in
clusters, their velocity distribution could in principle differ
significantly from that of the dark matter. Here we explore the
feasibility of using independent mass estimators to break degeneracies
in dynamical models.  We can then measure the orbital distribution of
cluster galaxies without assuming that cluster galaxies are good
tracers of the velocity field.

If as expected in the standard scenario, galaxies form primarily at
the centers of dark matter halos, the orbital distribution of cluster
galaxies is likely to be closely connected to that of the dark matter.
However, various physical mechanisms could produce a velocity bias
between the two components.  For instance, the formation of galaxy
pairs could transform orbital energy into internal energy and thus
produce velocity anti-bias (Fusco-Femiano \& Menci 1995).  Different
investigations with numerical simulations have produced a range of
predictions for the magnitude and even the sense of velocity bias
(Col{\'\i}n, Klypin \& Kravtsov ~2000, Ghigna et al. 2001, Diemand,
Moore \& Stadel 2005, Gao et al. 2005, Faltenbacher et al. 2005).
Col{\'\i}n et al.~(2000) find that cluster galaxies are positively
biased in the radial range (0.2-0.8)$r_{\rm vir}$, where $r_{\rm vir}$
is the virial radius, and approximately unbiased at larger radii.
They attribute this radial dependence to dynamical friction on cluster
galaxies near the cluster centre: dynamical friction is more efficient
at lower velocities, so these galaxies merge, and the remaining
galaxies are therefore preferentially high-velocity galaxies.
Faltenbacher et al. (2005) find that galaxies move slightly faster
than the dark matter particles with a velocity bias factor of
approximately 1.1.

Gravitational lensing offers perhaps the cleanest and most powerful
way to map mass distributions of galaxy clusters. The lensing mass
estimates can be used directly to solve for the orbits of galaxies in
the cluster. This approach was proposed and outlined in Natarajan \&
Kneib (1996) for clusters with mass estimates derived by combining
strong and weak lensing effects. Their proposed formalism can be
applied to any estimator of the mass profile that is independent of
the velocity structure and therefore be used to robustly constrain the
velocity anisotropy parameter.  One such estimator is X-ray
observations of the temperature and surface brightness distribution of
the intra-cluster medium (ICM). In this paper, using X-ray and lensing
mass profiles we study five clusters divided into two samples, three
clusters (A2199, A496 and A576) from the Cluster And Infall Region
Nearby Survey (CAIRNS; Rines et al.~2003) that constitute the nearby
sample and two clusters (A2390 and MS1358) that form our distant
sample
\footnote{Note that we use the word sample to define our chosen
clusters, which span a range of morphology, mass and X-ray
luminosity.}.  We solve for the nature of galaxy orbits via the
velocity anisotropy parameter $\beta_{\rm orb}(r)$, using the measured
galaxy velocities, the line-of-sight velocity dispersions and radial mass
profiles. Recently,
Mathews \& Brighenti (2003) in a similar vein used the X-ray
temperature profile of the elliptical galaxy NGC 4472 to constrain the
orbital distribution of its stars.
 
There are likely to be systematic uncertainties in both X-ray and
lensing mass estimates. For instance, the lensing signal around a
cluster may have contributions from associated large-scale structure,
and the overall normalization is typically unknown due to mass-sheet
degeneracy. The mass-sheet degeneracy can however be broken by the combined
use of strong and weak lensing features to constrain the mass profile
(Natarajan et al. 2002; Bradac, Lombardi \& Schneider 2004).
On the other hand, the intracluster gas may not be in equilibrium;
{\em Chandra} observations show evidence of buoyant bubbles near the
cores of clusters (McNamara et al.~2000; Fabian et al.~2000; Churazov
et al.~2001) as well as cold fronts and substructure (Markevitch et
al.~2000; Vikhlinin, Markevitch, \& Murray 2001). However, applying
these techniques to clusters has an advantage over ellipticals because
the hot gas is the dominant component of baryonic mass in clusters,
whereas it is a much smaller component of the baryonic mass in
ellipticals. For the clusters in our sample, the lensing mass
estimates have been obtained using strong and weak lensing data
enabling the calibration of the mass, therefore the mass-sheet 
degeneracy has been lifted (Natarajan et al. 2005).

The combined X-ray and optical data can constrain the properties of
both baryonic tracers - the galaxies and the gas. One of these
constraints comes from the so-called $\beta_X$
problem\footnote{Unfortunately, the traditional notation uses $\beta$
as the symbol for both the orbital anisotropy of tracers and the ratio
of energy per unit mass in galaxies and gas.  We use subscripts to
distinguish the two uses: $\beta_{\rm orb}$ refers to orbital
anisotropy and $\beta_X$ refers to the energy ratio.}.  The relative
internal energies of galaxies and gas in clusters can be characterized
by the parameter $\beta_{X} = {\mu m_p \sigma_r^2}/(k\,T_X)$ where
$\mu$ is the mean molecular weight, $m_p$ is the proton mass, $k$ is
Boltzmann's constant, $T_X$ is the X-ray temperature of the ICM, and
$\sigma_r^2$ is the radial velocity dispersion (usually estimated by
assuming isotropic orbits so that $\sigma_r^2$ equals the
line-of-sight velocity dispersion). The relative spatial distributions
of galaxies and gas can be characterized by a different parameter,
usually called $\beta_{gal}$, which should equal $\beta_{X}$ if the
populations are in equilibrium.

The ICM is assumed to be a fairly good tracer of the cluster
potential, despite the fact that it is unlikely to be in strict
hydro-static equilibrium. We are now able to test the robustness of
this assumption, as we can compare the measured temperature profile to
that predicted from the orbital model for the cluster. 

There is growing evidence that the ICM is not in thermal equilibrium,
in that instance it is unclear how to interpret $\beta_X$. Many studies
find an average value of $\beta_{X}\approx0.67$ (Jones \& Forman 1984,
1999).  Values of $\beta_{X}<1$ suggest that the ICM contains more
energy per unit mass than the galaxies.  This situation could arise
from energy input in the ICM or by some of the orbital energy of the
galaxies being converted into internal energies (Fusco-Femiano \&
Menci 1995).  We use either X-ray or lensing mass profiles to
determine the orbital distribution $\beta_{\rm orb}(r)$, which then
allows us to estimate $\sigma_r(r)$ and hence $\beta_{X}(r)$, the
energy ratio as a function of radius. Additionally, we compare the
results for our clusters with high-resolution cosmological N-body
simulations that include hydrodynamics (Faltenbacher et
al. 2005). These simulations trace the evolution of the gas and dark
matter and offer the ideal comparison set for our diverse clusters.

In this paper we demonstrate that with the wealth of multi-wavelength
data likely to be available in the very near future for large cluster
samples, fruitful comparisons can be made with samples of simulated
clusters selected in a similar fashion from high-resolution
cosmological N-body simulations. All the observational data to do so are
available at the present time only for a handful of clusters that
we study here.

The outline of our paper is as follows: in Section 2, we present the
formalism and the framework for our analysis.  In Section 3, we
discuss the derivation of the mass profiles from lensing and X-ray
observations. We discuss the galaxy velocity data in Section 4.  We
present the derived orbital distributions in Section 5.  Section 6
discusses the entropy profiles of the ICM.  We compare our
calculations with simulations in Section 7 and we conclude with a
discussion of the results in Section 8.  We assume a cosmology of
$\Omega_m=0.3$, $\Omega_\Lambda=0.7$, and $H_0=70$ km s$^{-1}$
Mpc$^{-1}$ throughout.

\section{The anisotropic Jeans equation}

The velocity anisotropy profiles of galaxy clusters can be obtained by solving 
the anisotropic Jeans equation, 

\be
\label{Jeans}
\frac{\ud (\nu_g \sigma_r^2)}{\ud r}
+ \frac{2 \beta_{\rm orb}(r) \nu_g \sigma_r^2}{r}
= - \frac{G M_{tot}(r) \nu_g}{r^2},
\ee

\noindent which applies to collisionless, spherically symmetric
systems of particles.  Here $M_{tot}(r)$ is the total cluster mass
contained within radius $r$ as determined by gravitational lensing or
X-ray observations, $\nu_g(r)$ is the three-dimensional galaxy number
density, $\sigma_r^2(r)$ is the radial velocity dispersion of the
galaxies, and $\beta_{\rm orb}(r)$ is the velocity
anisotropy\footnote{The reader should note that all quantities used in
this paper are three-dimensional, unprojected quantities, unless
explicitly stated otherwise.}. This parameter $\beta_{\rm orb}(r)$ is
defined to be $ (1 - \frac{\sigma_t^2}{\sigma_r^2})$ and is a measure
of the predominance of tangential orbits over radial ones. For
isotropic orbits $\beta_{\rm orb} = 0$; $0 < \beta_{\rm orb} < 1$ for
radial orbits and for tangential orbits $ -\infty < \beta_{\rm orb} <
0$.  The anisotropic Jeans equation above can be combined with the
equation that defines the measured line-of-sight velocity dispersion
$\sigma_{los}(R)$,

\begin{eqnarray}
\frac{1}{2}[\Sigma_g(R) \sigma_{los}^2(R)]
&=& \int_R^{\infty} \frac{r \nu_g(r) \sigma_r^2(r) \ud r}{\sqrt{r^2-R^2}}
\nonumber \\ \,\,\,\,\,\,\,\,\,\,\,\,&-& R^2 \int_R^{\infty} \frac{\beta_{\rm orb}(r) \sigma_r^2(r) \nu_g(r) \ud r}
{r \sqrt{r^2-R^2}},
\end{eqnarray}

\noindent to yield two independent integro-differential equations 
for $\sigma_r^2(r)$ and $\beta_{\rm orb}(r)$.
In the latter equation, $\Sigma_g(R)$ is simply the projected two-dimensional 
galaxy number density, which can be obtained by integrating $\nu_g(r)$. 
Following the procedure outlined in Natarajan \& Kneib 1996 (and in 
Bicknell et al. 1989), one can solve these equations to express $\sigma_r^2(r)$
as a sum of four integrals,

\be
\nu_g(r) \sigma_r^2 = {\rm Term_1(r)} - {\rm Term_2}(r) + {\rm Term_3}(r) - {\rm Term_4}(r), 
\ee

\be
{\rm Term_1}(r) = \frac{1}{3} \int_r^{R_t} \frac{G M_{tot}(r) \nu_g}{r^2} \ud r,
\ee

\be
{\rm Term_2}(r) = \frac{2}{3r^3} \int_0^r G M_{tot}(r) \nu_g r \ud r,
\ee

\be
{\rm Term_3}(r) = \frac{1}{r^3} \int_0^r R \Sigma_g(R) \sigma_{los}^2(R) \ud R,
\ee

\begin{eqnarray}
{\rm Term_4}(r) = \frac{2}{\pi r^3} \int_r^{R_t} R \Sigma_g(R) \sigma_{los}^2(R) 
\nonumber \\ \,\,\,\,\,\,\,\,\,\,\,\,\,
\Big( \frac{r}{\sqrt{R^2-r^2}} - \sin^{-1} \frac{r}{R} \Big) \ud R.
\end{eqnarray}

\noindent Here $R_t$ is a large radius at which both $\nu_g(r)$ and
$\Sigma_g(r)$ asymtote to zero. We find that using $R_t = 3.5$ Mpc
results in smooth $\beta_{\rm orb}(r)$ profiles in the range $0 -
1$ Mpc.  The above integrals can easily be computed numerically using a
constant step size trapezoid-rule algorithm. In the case of ${\rm
Term_4}$, a simple change of variable, $R = r \sqrt{1 + \zeta^2}$, is
used to eliminate the integrable singularity at the lower limit. After
computing $\sigma_r^2$, one can obtain $\beta_{\rm orb}(r)$ from the
Jeans equation,

\be
\beta_{\rm orb}(r) = - \frac{r}{2 \nu_g \sigma_r^2}
\Big[ \frac {G M_{tot}(r) \nu_g}{r^2} + \frac{\ud}{\ud r} (\nu_g \sigma_r^2)
\Big].
\ee

\section{Cluster Mass Profiles}

To solve the anisotropic Jeans equation above, we use the
independently determined mass profile for clusters $M(r)$ from lensing
and/or X-ray data.  The total mass profiles of all the clusters are
fitted with the NFW model. This `universal' profile provides the
best-fit to the structure of dark matter halos that form in N-body
simulations in the context of a cold dark matter cosmogony (Navarro,
Frenk \& White 1996). The NFW density profile can be written as

\be
\frac {\rho(r)}{\rho_{crit}} = \frac {\delta_c}{(r/r_s)(1+r/r_s)^2},
\ee

\noindent where $\rho(r)$ is the three-dimensional total density, 
$\rho_{crit} = 3 H^2 / 8 \pi G$ is the critical density ($H$ is the value of
the Hubble constant at the cluster's redshift), $r_s$ is the
characteristic scale radius, and 

\be
\delta_c = \frac {200}{3} \frac{c^3}{[\ln(1+c)-c/(1+c)]}.
\ee

The concentration parameter is defined by $c \equiv r_{200}/r_s$, where
$r_{200}$ is the radius within which the average density is 200 times
the critical density. The normalization, and hence the concentration
$c$, of the NFW model fits for our distant clusters is derived from the
detected strong lensing features at the Einstein radius.

\subsection{Mass profiles from gravitational lensing}

Strong and weak lensing data are used in combination to construct NFW
profiles for the distant clusters A2390 (Natarajan et al. 2005) and
MS1358 (Hoekstra et al.~1998). For MS1358 and A2390 the lensing mass
profiles are found to be in excellent agreement with the X-ray mass
profile (Arabadjis et al.  2002; Allen et al. 2001). In this work, we
use the NFW fits to the total mass for all 5 clusters.

\subsection{Mass profiles from X-ray data}

We use NFW mass profiles derived from X-ray observations for both the
nearby and distant sample clusters.  Table 1 presents the published fit
parameters and error bars for all five clusters.

\begin{table*}
\centering
\begin{minipage}{140mm}
\caption{Parameters for the total mass profile from X-ray data,
fitted to an NFW model with the reported errors in the fit parameters
and the consequent errors on the total mass. The errors reported are
1-$\sigma$ errors.}
\begin{tabular}{r c c c c l}
\hline
Cluster & $z$ & $r_s$, kpc & $c$ & $M$($r=0.5$ Mpc), $M_{\odot}$ &
Reference \\
\hline
A2199 & 0.0309 & 129$\pm$15 & 10 & $(1.33\pm0.15) \times 10^{14}$ & Markevitch et al. 1999 \\
A496 & 0.032 & 257$\pm$30 & 6 & $(1.66\pm0.18) \times 10^{14}$ & Markevitch et al. 1999 \\
A576 & 0.03829 & 300$\pm$40 & 4.5 & $(1.11\pm0.15) \times 10^{14}$ & Rines et al. 2000 \\
A2390 & 0.2284 & 629$^{+1090}_{-308}$ & 3.3 & $(3.26\pm0.98) \times 10^{14}$ & Allen et al. 2001\\
MS1358 & 0.3289 & 109$^{+116}_{-59}$ & 8.4 & $(1.42\pm0.28) \times 10^{14}$ & Arabadjis et al. 2002\\
\hline
\end{tabular}
\end{minipage}
\end{table*}

On large scales, the X-ray profiles of most of our clusters - A2199,
A496, A576 and MS1358 - appear smooth and circular, without any
indications of recent merging activity (Markevitch et al. 1999,
Kempner \& David 2004, Arabadjis et al. 2002). However, high-
resolution \textit{Chandra} observations reveal substructure in the core
regions of the nearby clusters.  A496 has a cold front with a sharply
decreasing surface brightness to the North (Dupke \& White 2003). A576
has surface brightness edges probably due to gas stripped off a small
merging subcluster in the centre (Kempner \& David 2004).  The ICM of
A2199 is interacting with lobes from the central radio source 3C 338
(Johnstone et al.~2002).  The X-ray image of A2390 looks smooth but
slightly elongated in different directions depending on radius. The
central region shows some substructure - a larger and a smaller X-ray
peak, suggesting that the cluster is not fully relaxed after recent
merger activity, although it appears to be in hydrostatic equilibrium
(Allen et al. 2001). 

The X-ray data provide estimates of the gas temperature profiles
needed to calculate the mass profiles and also $\beta_X(r)$ ($\S 6$).
Temperature profiles for the nearby clusters are available from both
\textit{Chandra} and \textit{ASCA/ROSAT} observations.  The
\textit{Chandra} data have better spatial resolution than the
\textit{ASCA/ROSAT} data, but extend to smaller radii (typically about
0.2 Mpc as opposed to approximately 1 Mpc for the \textit{ASCA} data).
\textit{Chandra} temperature data for A2199, A496 and A576 are
obtained from Johnstone et al.  (2002), Dupke \& White (2003), and
Kempner \& David (2004), respectively.  Johnstone et al. found the
temperature profile of A2199 to be spherically symmetric to a good
approximation. Consequently, they binned the data into concentric
annular rings, and fitted them using a single-temperature MEKAL model
for each annulus (see Johnstone et al. 2002 for details). We fit the
Johnstone temperature profile with a logarithmic function using a
least-squares algorithm. The best fit is: $ T = 1.657 \lg(r) + 1.035,
$ where $T$ is in keV and $r$ is in kpc.

Due to significant temperature anisotropy, Dupke \& White (2003)
divided A496 into two regions (SHARP and SMOOTH; see their Fig.~1b)
and obtained separate temperature profiles for each region using a
WABS MEKAL spectral model. We fit the temperature profiles with linear
functions using a least-squares algorithm. The best fit to the SHARP
data is $T = 0.0475 r + 1.75,$ where $T$ and $r$ are again in keV and
kpc, respectively.  The linear fit to the SMOOTH data, which is
somewhat less perfect, produces very similar results for
$\beta_X$. Therefore, we use the parameters for the SHARP data
hereafter.  Finally, Kempner \& David (2004) subdivided A576 into
three sectors (see their Fig. 2).  Their combined data are
consistent with a constant temperature of 4 keV for the entire range
of radii considered. Fig. 1 shows the temperature data and fits for
A2199, A496 and A576. All temperature profiles presented in this paper
are three-dimensional (deprojected) profiles. Markevitch et al.~(1998) 
showed that this model provides a good fit to an ensemble of 30 clusters 
observed with ASCA.  De Grandi \& Molendi (2002) found similar results
for the average cluster temperature profile of 21 clusters observed with 
BeppoSAX.  Because both of these satellites have poor angular resolution, 
the accuracy of these results has been controversial.  However, recent 
high-resolution {\em Chandra} measurements of temperature profiles in 11 
clusters confirm that cluster temperature profiles do decline in the 
range 0.2-0.6 $r_{180}$, although some clusters have "cooling cores" with 
smaller temperatures at radii $<$0.1$r_{180}$ (Vikhlinin et al.~2005).  
The apparent discrepancy between the {\em ASCA} and {\em Chandra} 
temperature profiles of A2199 and A496 in Figure 1 is thus most likely 
a result of the different spatial resolutions of the observations.  The 
{\em ASCA} temperature measurements are unreliable at very small radii 
due to poor spatial resolution, while the smaller field of view of {\em 
Chandra} prevents a direct comparison of the temperature profiles of 
these two clusters.  The agreement between the average temperature 
profiles of 11 clusters observed with {\em Chandra} (Vikhlinin et 
al.~2005) and 16 clusters observed with {\em XMM-Newton} (Piffaretti et 
al.~2005) with earlier ASCA and BeppoSAX results indicates that the 
ASCA temperature profiles are reliable at radii $>$0.1 Mpc.  At smaller
radii, the {\em Chandra} measurements are more reliable, although the 
complex properties of the ICM may indicate that the ICM is not in 
equilibrium at these radii.  The well-behaved declining temperature 
profiles at larger radii, however, are consistent with an ICM in 
approximate hydrostatic equilibrium.  Because we wish to use the ICM to 
estimate the mass profile (and because the radial range more closely 
matches that of the cluster galaxies), we use the ASCA temperature 
profiles of A2199 and A496 for our dynamical analysis.

\begin{figure}
\begin{center}
\includegraphics[height=9cm,width=9cm]{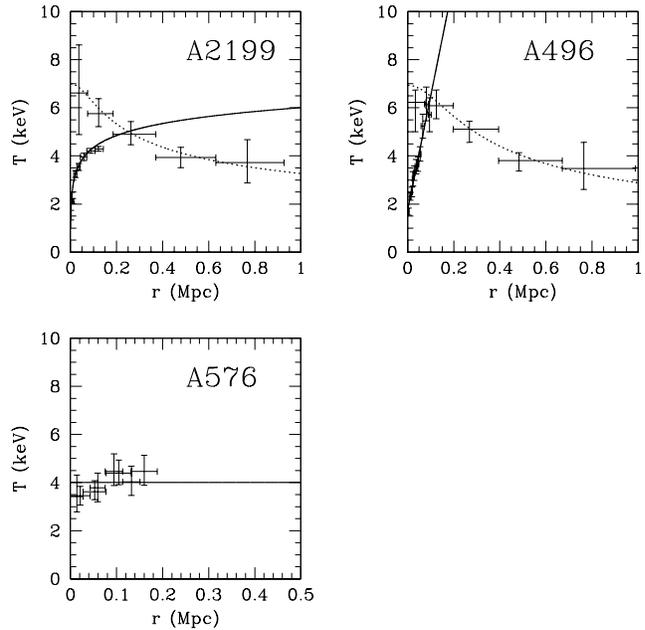}
\caption{Temperature profiles for the nearby clusters. For A2199 and A496, 
both the \textit{Chandra}
(solid line) and \textit{ASCA/ROSAT} (dotted line) temperature fits are
shown. The points are the observed data points. In the A576 plot 
the temperature data for two of the three cluster sectors (West and Southeast)
described in Kempner \& David 2004 are shown along with
our 4 keV fit line.}
\end{center}
\label{fig1}
\end{figure}

\begin{figure}
\begin{center}
\includegraphics[height=9cm,width=9cm]{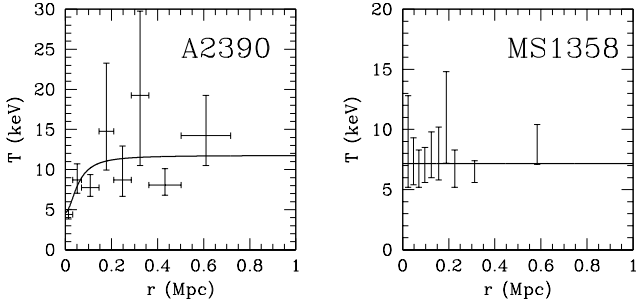}
\vspace{-4.5cm}
\caption{Temperature profiles from \textit{Chandra} data for the distant
clusters. The points and solid lines
indicate the observational data and fits, respectively.} 
\end{center}
\label{fig2}
\end{figure}

The \textit{ASCA/ROSAT} temperature data for A2199 and A496 are obtained from
Markevitch et al. (1999). The authors  fitted the data 
with polytropic models:

\be
T \propto \Big(1 + \frac{r^2}{r_X^2}\Big)^{-(3/2)\alpha (\gamma - 1)},
\ee

\noindent
where for A2199 $r_X = 95.7$ kpc, $\alpha = 0.636$, and $\gamma =
1.17$, and for A496 $r_X = 178$ kpc, $\alpha = 0.700$, and $\gamma =
1.24$.  The \textit{ASCA} data and fits for the nearby sample are shown
in Fig. 1. For A576, Rines et al. (2000) find a temperature of $3.77
\pm 0.10$ keV, and White (2000) finds $4.02 \pm 0.07$ keV, in good
agreement with the \textit{Chandra} value, suggesting that the cluster
is approximately isothermal.  The ASCA temperature profile of White
(2000) is consistent with isothermal to a radius of $15'= 0.9$Mpc.

The temperature profiles of the distant clusters are based on \textit{Chandra}
data. For A2390, Allen, Schmidt \& Fabian (2001) find that the
temperature is well described by:

\be
\frac {T(r)}{T_{2500}} = T_0 + T_1 \frac {(x/x_c)^{\eta}}{1+(x/x_c)^{\eta}},
\ee

\noindent where $x = r/r_{2500}$, $T_{2500} = 11.65^{+3.18}_{-2.45}$
keV, $r_{2500} = 0.64^{+0.15}_{-0.09}$ Mpc, $T_0 = 0.40 \pm 0.02$, $T_1
= 0.61 \pm 0.07$, $x_c = 0.087 \pm 0.011$, and $\eta = 1.9 \pm 0.4$.

The temperature profile of MS1358 is consistent with an isothermal
temperature of $T_X = 7.16 \pm 0.10$ keV (Arabadjis et al.~2002).
Fig. 2 shows the temperature data and fits for the distant sample.

\section{Velocity Dispersion Profiles}

The velocity data for our nearby clusters comes from the CAIRNS
project, which combines new redshifts with those from many earlier
studies.  Optical and X-ray mass estimates of A576 disagree (Mohr et
al.~1996, Rines et al.~2000), so it is of great interest to determine
if physically reasonable orbital distributions can reconcile this
conflict.  Markevitch et al.~(1999) proposed that A496 and A2199 are
good examples of relaxed clusters based on their circularly symmetric
X-ray contours and their well-behaved temperature profiles.  Despite
its quiescent central region, A2199 is surrounded by several infalling
X-ray groups and has an asymmetric galaxy distribution (Rines et
al.~2001, 2002).

We calculate velocity dispersion profiles (VDPs) for these three
clusters using galaxies defined to be members from the caustic
technique.  In redshift space, cluster infall regions exhibit dense
envelopes in redshift-radius diagrams.  These envelopes, known as
caustics, provide a straightforward membership classification.  The
use of caustics for defining cluster membership is similar to the
``shifting gapper'' technique (Fadda et al.~1996), but is less
sensitive to the density of the galaxy sample.  We use the caustics
found by Rines et al.~(2003) to define membership for the nearby
clusters. VDPs calculated with only galaxies brighter than $M_K=-23.5$
(Rines et al.~2004) in the Two-Micron All-Sky Survey (Nikolaev et
al.~2000) are consistent with those used here, suggesting that bright
and faint galaxies have similar orbital distributions and velocity
biases.  We calculate the velocity dispersion in bins of 25 galaxies;
changing the bin size does not significantly affect the VDPs.  The
typical redshift uncertainties are $\sim$30 km s$^{-1}$, so these
uncertainties contribute only a small amount of uncertainty to the
VDPs.  We do not attempt to subdivide the galaxy sample to obtain
different orbital solutions for different sub-samples (Biviano \&
Katgert 2004).

Like the total mass, the galaxy number density of the nearby clusters
is modeled with an NFW profile.  Rines et al.~(2004) calculate the
number density profiles of galaxies in A2199, A496 and A576 from
complete, $K_s$-selected spectroscopic samples.  The galaxy samples
are complete to 1.5-2 magnitudes fainter than $M_*$ and membership is
defined from spectroscopic redshifts using the caustic technique
(Diaferio 1999).  Rines et al.~(2004) fit the observed number density
profiles to NFW models and find scale radii of $r_{s,gxy}$=314 kpc,
500 kpc, and 1.07 Mpc for A496, A576, and A2199 respectively.  Because
these profiles are derived from complete, $K_s$-selected spectroscopic
samples, there is very little uncertainty in the profiles due to
incompleteness. The overall normalization of the galaxy density (and
thus the value of $c$) is immaterial since it cancels out in the
calculation of $\beta_{\rm orb}(r)$ (see Section 2). The total mass,
line-of-sight velocity dispersion and galaxy number density profiles
are all the ingredients needed to calculate $\beta_{\rm orb}(r)$, as
explained in Section 2.

For the two distant clusters, we again use the caustics to define
cluster membership (Diaferio et al.~2005).  Redshift data are
collected from the literature, with the majority coming from Yee et
al.~(1996) [A2390] and Fabricant et al.~(1991), Yee et al.~(1998), and
Fisher et al.~(1998) [MS1358].

The VDPs for all five clusters are fitted with polynomial functions
using a least-squares algorithm.  The lowest degree polynomial
producing a good fit (typically $\chi^2/DOF < 1$) is used. The
polynomial fits are applied in the radial range $0 < r < r_{max}$,
where $r_{max}$ is the maximum radius for which VDP data are
available. For $r_{max} < r < 5$ Mpc, $\sigma_{los}$ is set equal to
the value at $r_{max}$, and beyond that it is set equal to zero. We
exclude the innermost data points for A496 and A2390 because including
them requires higher-order polynomials to yield acceptable fits to the
VDP data; these polynomials lead to unphysical solutions for
$\beta_{orb}(r)$.  Figure 3 shows the VDP data and fits for the five
clusters.

\begin{figure}
\begin{center}
\includegraphics[height=9cm,width=9cm]{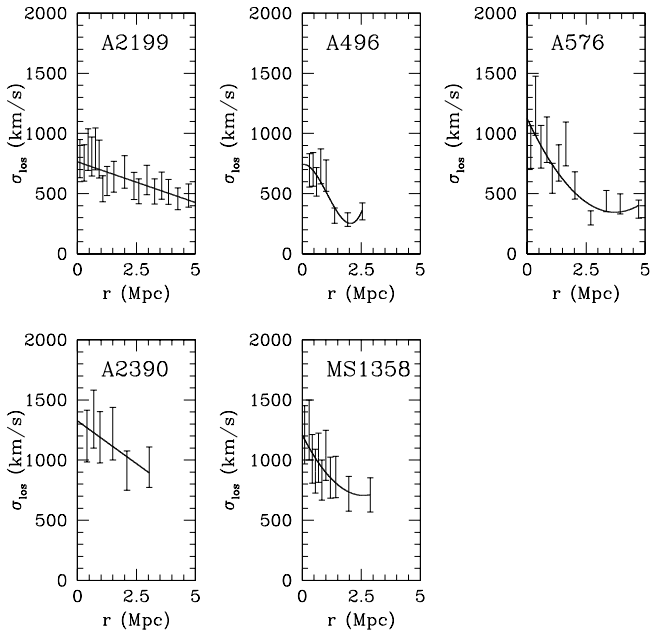}
\caption{Velocity dispersion profiles and polynomial fits for all five
clusters.}
\end{center}
\end{figure}

The galaxy number densities of A2390 and MS1358 are well-described by
a Hernquist profile:

\be
\nu_g (r) \propto \frac{1}{(r/r_c)(1+r/r_c)^3}
\ee

\noindent 
with $r_c = 1.42$ Mpc and $r_c = 1.08$ Mpc, respectively (Carlberg et
al.  1997).
 
\begin{figure}
\begin{center}
\includegraphics[height=9cm,width=9cm]{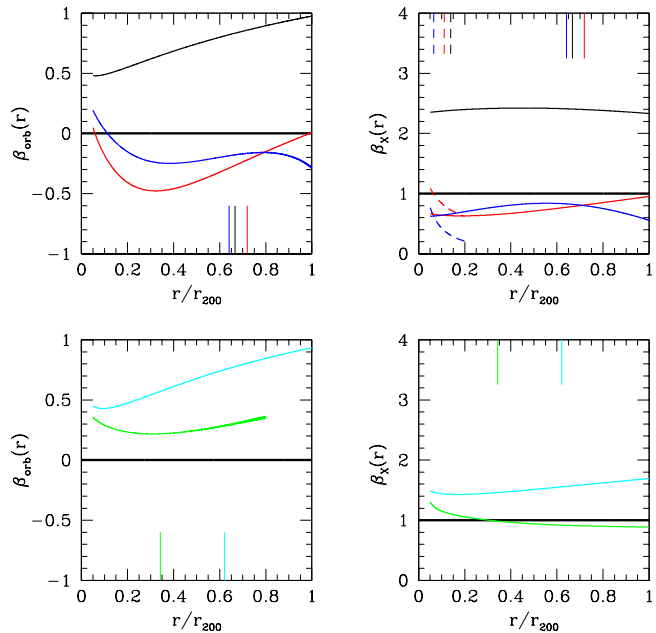}
\caption{Velocity anisotropy parameter $\beta_{\rm orb}$ and 
energy ratio $\beta_X$. 
The top-left panel shows $\beta_{\rm orb}$ for the nearby clusters: 
A2199 (red curve), A496 (blue curve) and A576 (black curve). 
The top-right panel shows $\beta_X$ for the nearby clusters computed
using \textit{ASCA/ROSAT} and \textit{Chandra} temperature data. Colors are
as for the left panel: solid red - A2199 (\textit{ASCA/ROSAT}), 
solid blue - A496 (\textit{ASCA/ROSAT}), 
black - A576 (\textit{ASCA} and \textit{Chandra}), 
red dashed - A2199 (\textit{Chandra}) 
and blue dashed - A496 
(\textit{Chandra}). The bottom-left panel shows $\beta_{\rm orb}$ for the
high-redshift clusters. Lines are as follows: green curve - MS1358 and cyan 
curve - A2390.
Finally, the bottom-right panel shows $\beta_X$ for the distant clusters
(lines are as in the bottom-left panel). The vertical ticks show the maximum
extent of the X-ray temperature data for the left panels. For the
right panel the vertical ticks show the extent to which the mass profile
is fitted. The thick line marks $\beta_{\rm orb} = 0$ and 
$\beta_X = 1$.}
\end{center}
\end{figure}

The CNOC1 group estimate the galaxy number density profiles of A2390
and MS1358 from their large spectroscopic samples including a
completeness correction for galaxies without redshifts which depends
on both position and magnitude (Carlberg et al.~1997).  They find that
the number density profiles are well fit by a Hernquist profile: [eqn.~
13] with $r_{s,gxy}$=1.42 and 1.08 Mpc for A2390 and MS1358
respectively.

\section{Results for the velocity anisotropy parameter}

The results for $\beta_{\rm orb}$ for all five clusters are shown in
the left-hand panels of Fig. 4. We find diversity in orbital
parameters in our sample. Two of the clusters A2199 and A496 in the nearby sample appear to
have predominantly tangential orbits in the inner region (out to 1
Mpc) as indicated by the behaviour of $\beta_{\rm orb}(r)$, whereas A576 appears
to have mostly radial orbits. The
distant clusters seem to be dominated by radial orbits. 
Computing $\beta_X$ using the observed
temperature profile and the $\sigma_r^2(r)$ computed from the
anisotropic Jeans equation, we find that MS1358 and A576 appear to
deviate from hydro-static equilibrium ($\beta_X \ne 1$) from the inner 
regions to all the way out, and the rest 
seem to be consistent with being by-and large in hydro-static equilibrium.
For A576 and MS1358, $\beta_X > 1$ indicating less energy in
the gas compared to the galaxies, whereas for A2199, A496 and A2390
$\beta_X \leq 1$. For A2199 and A496, the $\beta_X$ profiles computed
from the \textit{ASCA/ROSAT} temperature data are slightly higher than
those from the \textit{Chandra} data and have a positive
slope. Henceforth, we focus on the former because of the larger
spatial extent necessary for comparison with the numerical
simulations. We show in Figure 5 the effect of changing the
normalization of the total VDPs by $\pm$10\%.

\section{Comparison with cosmological simulations}

We now compare the observational results for $\beta_{\rm orb}$,
$\beta_X$ and $S(r)$ with those of high-resolution cosmological simulations based
on the ``concordance'' flat $\Lambda\mathrm{CDM}$ model: $\Omega_m = 1
- \Omega_{\Lambda} = 0.3$, $\Omega_b = 0.021$ $h^{-2}$, $h = 0.7$ and
$\sigma_8 = 0.9$, where the present-epoch Hubble constant is defined
as $100$ $h$ km $\mathrm{s}^{-1}$ $\mathrm{Mpc}^{-1}$, and $\sigma_8$
is the power spectrum normalization on $8$ $h^{-1}$ Mpc scale. The
simulations are performed using the Adaptive Refinement Tree (ART)
N-body+gasdynamics code (Kravtsov, Klypin \& Khokhlov 1997; Kravtsov
1999; Kravtsov, Klypin \& Hoffman 2002), an Eulerian code with
adaptive refinement in space and time and non-adaptive refinement in
mass.

To set up initial conditions we first run a low-resolution simulation
of $80$ $h^{-1}$~Mpc and $120$ $h^{-1}$~Mpc boxes and select six clusters
with virial masses ranging from $\approx 6\times10^{14}$ to
$1.6\times 10^{15}$ $h^{-1}{\ \rm M_{\odot}}$. The
largest one has a virial mass of $1.6 \times 10^{15}$ $h^{-1}$
$M_{\odot}$, the second massive cluster has a mass of $1.1 \times
10^{15}$ $h^{-1}$ $M_{\odot}$, and the other four clusters have masses
of $\approx 7-10 \times 10^{14}$ $h^{-1}$ $M_{\odot}$. The corresponding
virial radii, defined as radii enclosing the overdensity of 337 with
respect to the mean density of the Universe, for the six clusters are:
$2.36$, $2.10$, $2.01$, $1.93$, $1.92$, and $1.78$~$h^{-1}$~Mpc.

The initial conditions are set using multiple-mass particle technique
retaining the previous large-scale waves intact but including
additional small-scale waves, as described by Klypin et al. (2001).
The re-sampled lagrangian region of each cluster, corresponding to a
sphere of $(1.5-5)R_{\rm vir}$ around it at $z=0$, is then simulated
with high dynamic range.  The actual spatial resolution of the
simulations is $\approx 5$ $h^{-1}$ kpc. The mass resolution (i.e., the
dark matter particle mass) is $m_p\approx 9.1\times 10^8$ $h^{-1}$
$M_{\odot}$ for the all but the most massive cluster and $2.7\times
10^8$ $h^{-1}$ $M_{\odot}$ for the most massive cluster. The
clusters thus have million or more dark matter particles within their
virial radii. The
simulations follow dissipationless dynamics of dark matter particles
and gasdynamics of the baryonic component and include several
processes crucial to galaxy formation: star formation, metal
enrichment and thermal feedback due to supernovae types II and Ia,
self-consistent advection of metals, metallicity-dependent radiative
cooling and UV heating due to cosmological ionizing background.
Stellar feedback on the surrounding gas includes injection of energy
and metals via stellar winds and supernovae as well as secular mass
loss (see Kravtsov, Nagai \& Vikhlinin 2005, for details).

For each cluster simulation, we identify the main cluster and its
galaxies using a variant of the Bound Density Maxima (BDM) algorithm
using dark matter particles only.  The details of the algorithm and
parameters used in the halo finder can be found in Kravtsov et
al. (2004). The main steps of the algorithm are identification of
local density peaks (potential halo centers) and analysis of the
density distribution and velocities of the surrounding particles to
test whether a given peak corresponds to a gravitationally bound
clump. More specifically, we construct density, circular velocity, and
velocity dispersion profiles around each center and iteratively remove
unbound particles using the procedure outlined in Klypin et
al. (1999).  We then construct final profiles using only bound
particles and use these profiles to calculate properties of halos,
such as the circular velocity profile $V_{\rm
circ}(r)=\sqrt{GM(<r)/r}$ and compute the maximum circular velocity
$V_{\rm max}$.  For halos located within the virial radius of a larger
host halo ({\it the subhalos}), we define the outer boundary at the
{\rm truncation radius}, $r_{\rm t}$, at which the logarithmic slope
of the density profile constructed from the bound particles becomes
larger than $-0.5$ as we do not expect the density profile of the CDM
halos to be flatter than this slope. For each system we estimate the
stellar mass, $M_{\ast}$, gas mass and total mass (dark matter, stars,
and gas) within the truncation radius.

Once the galaxies are identified, the radial and tangential components
of the velocity dispersion, $\sigma_r$ and $\sigma_t$, of the dark
matter, gas and galaxies are measured in radial bins centered on the
cluster potential minimum after subtracting the peculiar velocity of
the cluster, defined as the mass-weighted bulk velocity of dark matter
within the cluster core. Only galaxies with masses $M_{\ast}> 1 \times
10^9$ $M_{\odot}$ (or $V_{\rm max}> 80$ $\rm km\,s^{-1}$) are used in
this calculation.  This value for the threshold provides a reasonably
large sample of galaxies, while not compromising the numerical
resolution of the results. The velocity anisotropy parameter
$\beta_{\rm orb}$ is computed from the components of the velocity
dispersion using the definition in Section 2. In addition, $\beta_X$
is computed using the galaxy $\sigma_r^2$ and the X-ray temperature of
the gas. The measured temperatures are gas mass weighted is calculated
assuming \textit{Chandra} energy response in the 0.5-7 keV band. The
temperature profiles in the simulations are extracted from the 3D
distribution and have negligible measurement errors.

Unfortunately, the numerical results are rather noisy in the central
cluster regions due to the smaller number of galaxies there.  To allow
a better comparison with the numerical simulations, we extend the
observed $\beta_{\rm orb}$ and $\beta_X$ profiles to 2 Mpc by
increasing the integration limit $R_t$ from 3.5 to 7 Mpc. Beyond
1.5 Mpc, there is a small amount of scatter in the $\beta_{\rm orb}$
profile due to numerical effects. Increasing $R_t$ beyond 7 Mpc
slows down the calculation considerably without reducing this scatter
much further, so 7 Mpc is established to be the optimal value for
the large $R$ limit for the purposes of this calculation.

\begin{figure}
\begin{center}
\includegraphics[height=9cm,width=9cm]{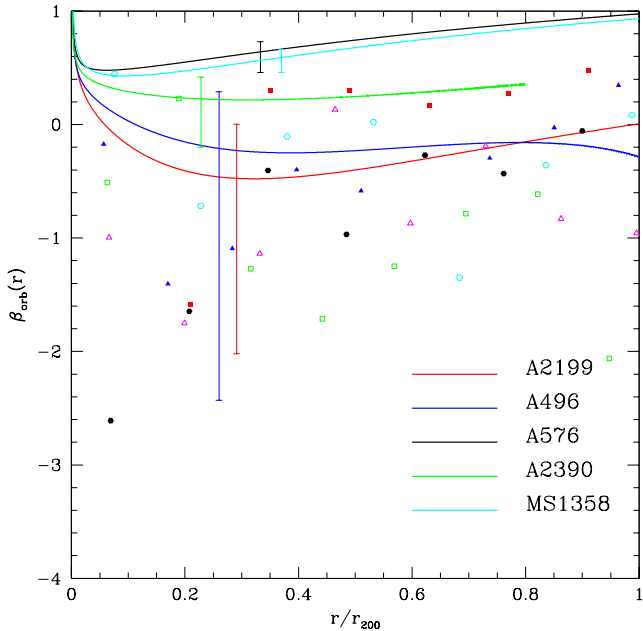}
\caption{Velocity anisotropy $\beta_{\rm orb}$ - comparison between
 observations and numerical simulations. The error bars on each curve
 shows the error in $\beta_{\rm orb}$ due to changing $\sigma_{los}$ by
 $\pm 10 \% $, evaluated at the given radius. The points represent the
 six simulated clusters.}
\end{center}
\end{figure}

\begin{figure}
\begin{center}
\includegraphics[height=9cm,width=9cm]{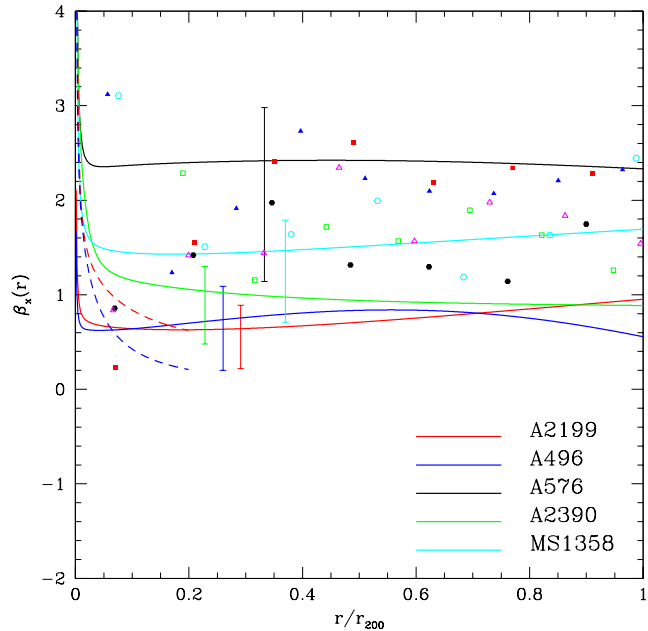}
\caption{Energy ratio $\beta_X$ - comparison between observations and
 numerical simulations. The $\beta_X$'s are computed from the
 $\beta_{\rm orb}$ using \textit{ASCA/ROSAT} temperature data for the
 nearby clusters and \textit{Chandra} data for the distant ones.  The
 error bars on each line show the error in $\beta_X$ due to changing
 $\sigma_{los}$ by $\pm 10 \% $, evaluated at the given radius.  The
 points represent the six simulated clusters.}
\end{center}
\end{figure}

The detailed results of the comparison with the simulations are discussed in
the section below. 

\section{Discussion and Conclusions } 

The statistical uncertainties in the individual data 
points for the VDPs and the number density profiles are $\sim$20\%; the 
uncertainties in the mass profiles are $\sim$10\%.  To illustrate the 
impact these uncertainties have on the derived galaxy orbits, we 
recalculate the orbital distribution $\beta_{orb}(r)$ after 
renormalizing the entire VDP by $\pm$10\%  (the uncertainty in the 
total VDP is smaller than the individual bins).  Because the 
$\beta_{orb}(r)$ profiles are much more sensitive to changes in VDP 
than either $\nu(r)$ or $M(r)$, these uncertainties are representative
of the total uncertainty from the observations (numerical errors are 
significantly smaller than the observational uncertainties). Figures 
5 and 6 show the magnitude of these uncertainties; they are comparable 
to the scatter seen in the simulated clusters.  The derived orbital 
profiles agree well with the simulated clusters, although the 
uncertainties in both observations and simulations are still large 
enough to make it difficult to distinguish mildly radial orbits from 
mildly circular orbits.  As the observations and simulations improve, 
more detailed treatments of the systematic uncertainties will be 
required.  The current study demonstrates the feasibility of this 
technique and the consistency of various mass estimators.  That is, 
we successfully derive physical orbit models for the cluster galaxies 
assuming that they trace the mass profile inferred from independent 
X-ray or lensing data.  We further show that these orbital profiles 
closely resemble those of cluster galaxies in simulations

In our cluster sample, we find a diversity of galaxy orbits. All
nearby clusters appear to have primarily tangential orbits.  The
distant cluster MS1358 exhibits radial orbits while A2390
has predominantly tangential orbits at small radii but radial
orbits at large radii ($r > 1$ Mpc). The
statistical uncertainties in the observed quantities can lead to large
uncertainties in $\beta_{\rm orb}$: the statistical $\approx 1 \sigma$
uncertainties are $\approx 20 \%$ for the VDPs, $\approx 10 \%$ for
the mass profiles, and $\approx 20 \%$ for the galaxy number
densities. As an illustration, the errors plotted in Fig. 6 and Fig. 7 are calculated by
changing $\sigma_{los}$ by $\pm 10 \%$, while keeping the total mass
and galaxy number density fixed.  Because $\beta_{\rm orb}$ is much
more sensitive to changes in the VDP than to the other two
quantities, these errors are representative of the overall
uncertainty in $\beta_{\rm orb}$ and $\beta_X$. Numerical errors are
generally within a few per cent and much smaller than the
observational uncertainties. At larger radii, where observational
constraints (lensing and X-ray data) need to be extrapolated, $\beta_{\rm
orb}$ is not directly constrained by the data. 

We find departures from hydrostatic equilibrium in all of our clusters
except A2390 (especially when the X-ray mass is used to compute
$\beta_{\rm orb}$). However, these deviations are generally small
($0.6 < \beta_X < 1.6$ excluding the \textit{Chandra}-based results
for A2199 and A496) and comparable to the margins of error (see
Fig. 6).  This implies that the cluster mass estimates from X-ray
observations, which are based on the assumption of hydrostatic
equilibrium, may still be valid.  A2199, A496 and A2390 have $\beta_X
< 1$, indicating more energy in the ICM than in the galaxies, whereas
A576 and MS1358 have $\beta_X > 1$, indicating less energy in the gas.

There is general agreement in shape, magnitude and range between the
observed and simulated $\beta_{\rm orb}$ profiles (Fig. 5), especially
for $r > 0.5$ Mpc, where the simulations are reliable due to
sufficient galaxy statistics. This agreement implies that the
anisotropic Jeans equation with its assumptions of spherical symmetry
is applicable to observational data. The simulated $\beta_X$ profiles also have similar shapes
and ranges to the observed ones, but their values are
approximately a factor of two larger (Fig. 6). The reason for this
discrepancy is not clear. Our results are consistent with 
theoretical expectations that clusters are dynamically complex as they are 
still in the process of assembling.   The difference between the inferred
orbits is curious because it suggests that the moderate redshift clusters 
are dynamically younger than the nearby clusters.  Due to the small sample 
size, however, it would be premature to conclude that cluster galaxy orbits 
evolve dramatically between $z$=0.2-0.3 and the present.  If this effect is
real, it could result from the observed evolution in the infall rate of 
galaxies (Ellingson et al.~2001).

There are several directions in which the present work could be
extended. First, one could use a larger cluster sample to verify
the statistical significance of the results presented in this paper - 
e.g. whether the differences in orbital structure between the nearby and
distant clusters are a coincidence due to our small samples or a
more general fact. In addition, better observational data, especially 
more precise cluster VDPs, would constrain the velocity anisotropy 
much further. The numerical simulations could also be refined 
in order to resolve better the central cluster regions. New sources of 
entropy injection could be explored with simulations that
include processes such as heating by quasars and AGNs, thermal conduction 
and turbulent mixing.
 
\section*{Acknowledgments} 

This research has made use of the NASA/IPAC Extragalactic Database
(NED) which is operated by the Jet Propulsion Laboratory, California
Institute of Technology, under contract with the National Aeronautics
and Space Administration. PN acknowledges financial support from
proposals HST-AR-10302.01-A and HST-GO-09722.06-A provided by NASA
through a grant from STScI, which is operated by AURA.

\end{document}